\documentclass[twoside,twocolumn]{article}

\usepackage{blindtext} 

\usepackage[sc]{mathpazo} 
\usepackage[T1]{fontenc} 
\usepackage{microtype} 

\usepackage[english]{babel} 

\usepackage[hmarginratio=1:1,top=32mm,columnsep=20pt]{geometry} 
\usepackage[hang, small,labelfont=bf,up,textfont=it,up]{caption} 
\usepackage{booktabs} 

\usepackage{lettrine} 

\usepackage{enumitem} 

\usepackage{abstract} 

\usepackage{titlesec} 
\renewcommand\thesection{\Roman{section}} 
\renewcommand\thesubsection{\roman{subsection}} 
\titleformat{\section}[block]{\large\scshape\centering}{\thesection.}{1em}{} 
\titleformat{\subsection}[block]{\large}{\thesubsection.}{1em}{} 

\usepackage{fancyhdr} 
\usepackage{titling} 

\usepackage{hyperref} 
\usepackage{natbib}
\usepackage[pdftex]{graphicx}     

\pretitle{\begin{center}\Huge\bfseries} 
\posttitle{\end{center}} 
\title{New Perspectives on Frontal Variability in the Southern Ocean Using a Local Identification Scheme} 
\author{%
\textsc{Christopher C. Chapman}\thanks{\textit{Corresponding author address:} 
				C. C. Chapman, LOCEAN-IPSL, 
				Universit\'{e} de Pierre et Marie Curie, Paris CEDEX ,France. 
				\newline{E-mail: chris.chapman.28@gmail.com}} \\[1ex] 
\normalsize LOCEAN-IPSL\\ Universit\'{e} de Pierre et Marie Curie \\ 
\normalsize \href{mailto:chris.chapman.28@gmail.com}{mailto:chris.chapman.28@gmail.com} 
}
\date{\today} 
\renewcommand{\maketitlehookd}{%
\begin{abstract}
\noindent The frontal structure of the Southern Ocean is investigated using a sophisticated frontal detection methodology, the Wavelet/Higher Order Statistics Enhancement (WHOSE) method, introduced in \cite{Chapman2014}. This methodology is applied to 21 years of daily gridded sea-surface height (SSH) data to obtain daily maps of the locations of the fronts. By forming `heat-maps' of the frontal occurrence frequency and then approximating these heat-maps by a superposition of simple functions, the time-mean locations of the fronts, as well as a measure of their capacity to meander, are obtained and related to the frontal locations found by previous studies.  

The spatial and temporal variability of the frontal structure is then considered. The number of fronts is found to be highly variable throughout the Southern Ocean, increasing (`splitting') downstream of large bathymetric features and decreasing (`merging') in regions where the fronts are tightly controlled by the underlying topography. In contrast, frontal meandering remains relatively constant. Contrary to many previous studies, little no southward migration of the fronts over the 1993-2014 time period is found, and there is only weak sensitivity to atmospheric forcing related to SAM or ENSO. The reasons for the discrepancy between this study and previous studies using contour methods are investigated and it is shown that the spatial variability of the frontal structure is not tied to the underlying sea-surface height. It is argued that the results of studies using sea-surface height contours to define front must be interpreted with care.  
\end{abstract}
}

\begin{document}

\maketitle


\section{Introduction}

\lettrine[nindent=0em,lines=3]{O}bservations dating back to Discovery expedition have revealed that the Antarctic Circumpolar Current (ACC) is composed of a number of large-scale hydrographic ``fronts" \citep{Deacon1937}. Although it is difficult to give a precise definition of a front \citep{LanglaisEtAl2011,Chapman2014}, they are generally considered to be the boundaries between different watermasses \citep{Sokolov&Rintoul2002}. In the Southern Ocean, fronts are aligned more-or-less in the zonal direction. Water mass properties change rapidly across the front, yet remain approximately constant along the front \citep{Deacon1937,OrsiEtAl1995,Belkin&Gordon1996}. These studies have lead to the ``traditional"  view \citep{LanglaisEtAl2011} that the Southern Ocean is composed of three circumpolar hydrographic fronts which are (from north to south) the Subantarctic Front (SAF), the Polar Front (PF) and the Southern ACC Front (sACCF). 

Southern Ocean fronts are thought to be closely related to strong zonal geostrophic jets \citep{Gille1994,Sokolov&Rintoul2007}, and although the details of the relationship between the hydrographic fronts and jets is still unclear \citep{GrahamEtAl2012, Chapman2014}, the terms ``jet" and ``front" are often used interchangeably. To further complicate the picture, there is evidence from high resolution satellite data and numerical modelling that these jets are not smooth, continuous circumpolar features, but are instead complicated meso-scale phenomena that split, merge and drift, and that their structure varies in space and time \citep{Hughes&Ash2001,Thompson2010,ThompsonEtAl2010,HughesEtAl2010,Hallberg&Gnanadesikan2006,Chapman2014}. 
Southern Ocean fronts  can influence the upwelling and ultimate ventilation of deep waters, which can in turn affect the formation of watermasses at the surface \citep{BoeningEtAl2008,MeijersEtAl2012,Meijers2014} and can act to suppressed meridional exchange of tracers \citep{FerrariandNikurashin2010,Thompson&Sallee2012}. Frontal anomalies could also result in anomalous water mass properties being redistributed throughout the ocean by the strong zonal currents that compose the ACC \citep{SalleeEtAl2008}. 

Multiple studies using either satellite data alone \citep{SokolovAndRintoul2009II,BillanyEtAl2010}, or mix of satellite and hydrographic data \citep{SalleeEtAl2008,Tarakanov&Gritsenko2014,Kim&Orsi2014} have identified large meridional shifts in the locations of fronts, sometimes as large as 10$^{\circ}$ of latitude. \cite{SokolovAndRintoul2009II} have found a 60km southward shift in the location of all major frontal branchs over the 15 years of altimetry data used in their study, and that temporal variability is concentrated in regions away from large bathymetric features. \citet{Kim&Orsi2014}, using similar methods, find that the magnitude of the southward shift in the front positions is spatially variable, being strongest in the southeast Indian basin, followed by the southwest Atlantic basin and weakest in the south Pacific. Both \cite{SalleeEtAl2008} and \cite{Kim&Orsi2014} find significant correlation between the Southern Annular Mode and the El--Ni\~{n}o Southern Oscillation and the position of the fronts in the southeast Indian and southeast Pacific basin.   	


In contrast, other studies using similar data yet different methods yield contrary results, showing no coherent trends in frontal location \citep{GrahamEtAl2012,Gille2014,ShaoEtAl2015,FreemanEtAl2016}, although these studies do not rule out localized or higher-frequency variability such as that meridional ``drift" of jets identified by \citet{Thompson&Richards2011} or ``jet-jumping" \citep{Chapman&Hogg2013,Chapman&Morrow2014a}. \citet{GrahamEtAl2012} find in a high-resolution numerical model that the location of the ACC fronts are insensitive to relatively large changes in the wind forcing, even over abyssal plains with limited influence from topography.  Using satellite alimetry and a frontal definition based on local skewness minima \cite{ShaoEtAl2015} found little evidence of frontal migration or sensitivity of frontal positions to atmospheric forcing. \cite{Gille2014} used zonally averaged ACC transport to compute an effective `mean latitude' of the ACC also found no evidence of any shift in the fronts, while \cite{FreemanEtAl2016} found that the location of the polar-front, as defined by sea-surface temperature gradients, had also shown limited temporal variability over the period 2002-2014.        

The root of this controversy appears to stem from the definition of a front. The ``contour methods", introduced in \cite{Sokolov&Rintoul2002} and used in the majority of studies that find long-term trends in frontal positions, exploit the fact that regions of high sea-surface height (SSH) gradient (that define a strong geostrophic current) are often collocated with a unique value of SSH. Given a robust trend in SSH has been detected \citep{MorrowEtAl20008}, it follows that if fronts are linked to a particular SSH value, any shift in the location of that contour will logically correspond to a shift in the front. In contrast, the majority of studies that find little or no change in frontal position define fronts based on \textit{local} criteria such as SSH gradients \citep{GrahamEtAl2012,Gille2014,ShaoEtAl2015}. Contour methods have recently been criticized by \cite{ThompsonEtAl2010} and \cite{GrahamEtAl2012}, who show that the unique value of SSH selected to designate a front may not always coincide with the regions of high SSH gradients, which can result in spurious temporal variability. Additionally, as noted by \citet{LanglaisEtAl2011}, when using contour type definitions, there is no agreed upon number of fronts in the Southern Ocean. A detailed discussion of the differences between frontal definitions and the implications of choosing one over another can be found in \citet{Chapman2014}. 

Although they have been the subject of criticism, contour methods have numerous advantages. In particular, contour methods provide an unambiguous frontal location at all longitudes. In contrast, local definitions are not so easy to work with. Fronts defined locally spontaneously form and dissipate; split or merge; and meander. As such, the number of fronts, and their latitude, varies with both longitude and time, which makes robust calculations of trends difficult to obtain. Effective treatment of an arbitrary number of interleaving fronts has not yet been accomplished.   

The goals of this paper are to study in detail the spatial and temporal variability of the frontal structure of the Southern Ocean and to investigate the response of fronts to changes in atmospheric forcing associated with the SAM and ENSO, while avoiding the shortcomings of contour methods. To do this, we will apply a sophisticated jet detection methodology, the Wavelet/Higher Order Statistics Enhancement (WHOSE), introduced in \cite{Chapman2014}, to 21 years of altimeteric sea-surface dynamic topography data. By approximating the `heat maps' of frontal occurrence frequency as a superposition of simple functions to avoid tracking individual frontal filaments, we will analyze how the frontal structure changes geographically and robustly determine trends in the frontal location, making no assumptions regarding the number of fronts or their circumpolar extent. We will also attempt to shed light on the reasons for the discrepancies between contour methods and local methods for detection fronts, and make some concrete recommendations concerning their use.

The paper is organized as follows: section \ref{Section:Data_and_Methods} will briefly discuss the frontal detection methodology (the WHOSE method) and describe the data to be used. Section \ref{Section:Mean_Front_Positions} will focus on the time mean frontal structure and its spatial variability. Trends in frontal locations and their sensitivity of changes in atmospheric forcing will be 	presented in Section \ref{Section:Variability_In_Front_Positions}. Finally, section \ref{Conclusion} discusses the implications of this work and makes a number of recommendations for future studies pursuing this subject.


\section{Data and Methodology} \label{Section:Data_and_Methods}

\subsection{Data}
\subsubsection{Time--mean dynamic topography}
To estimate the mean state of the ocean we use the combined mean dynamic topography (MDT) product, described in \cite{RioEtAl2014} and downloaded from CNES-CLS13 (http://www.aviso.altimetry.fr/fr/donnees/produits/produits-auxiliaires/mdt.html). The mean sea surface is reconstructed over the period 1993--2012 by combining data from the Gravity Recovery and Climate Experiment (GRACE) mission, satellite altimeters, and drifting buoys.  

\subsubsection{Time--varying dynamic topography}
The satellite dataset used in this study is the Archiving, Validation, and Interpretation of Satellite Oceanographic data (AVISO) daily gridded sea level anomalies (SLA) from Ssalto/Duacs, downloaded from Copernicus Marine Services (http://marine.copernicus.eu/web/69-interactive-catalogue.php) \citep{PujolEtAl2016}. We use delayed-mode dynamic topography for the period 1993--2015, with daily output, giving 8035 data records. These data are mapped to a 1/4$^{\circ}$ Mercator grid using optimal interpolation of alongtrack data series based on the REF dataset, which uses two satellite missions [Ocean Topography Experiment(TOPEX)/Poseidon/European Remote Sensing Satellite(ERS) or Jason-1/Envisat or Jason-2/Envisat] with consistent sampling over the 21-yr period. Data are corrected for instrumental errors, atmospheric perturbations, orbit errors, tides, inverted barometer bias, and aliased fast barotropic signals (periods of less than 20 days). Although the output grid spacing is 1/4$^{\circ}$, the effective resolution is, in fact, much lower \citep{PujolEtAl2016}. 

This study uses ``absolute dynamic topography" (ADT) which is the sum of the time-mean dynamic topography and time-varying sea level anomalies.
	
\subsection{Southern Annular Mode and El Ni\~{n}o Southern Oscillation Indies}
	
The leading modes of atmospheric variability over the Southern Ocean are generally considered to be the Southern Annual Mode (SAM) and the El Ni\~{n}o Southern Oscillation. To investigate the response of Southern Ocean fronts to changes in these climate modes, we obtained the monthly SAM index, described in \cite{Marshall2003}, from the British Antarctic Survey (http://www.nerc-bas.ac.uk/icd/gjma/sam.html) and the monthly Bivariate ENSO Timeseries \citep{Smith&Sardeshmukh2000} from the NOAA Earth Systems Research Laboratory (http://www.esrl.noaa.gov/psd/people/cathy.smith/best/). Both timeseries are then low-pass filtered with a finite-impulse response Blackman filter with a cut-off period of 3 months. 

\subsection{Front Detection Methodology}
Here we will briefly describe the Wavelet/Higher Order Statistics Enhancement (WHOSE) methodology used to detect fronts in this study. The WHOSE method was originally developed by \cite{Chapman2014}, inspired by methods used for the detection of step-like signals in acoustic and radar data. Further details, including implementation notes, extensive validation and a detailed comparison with other methods, are found in \cite{Chapman2014}. 

The WHOSE method belongs to the broader class of local gradient maxima methods. As their name implies, these methods identify fronts by finding points where the gradient of SSH, SST or some other quantity, exceeds a predefined threshold. Studies using this methods have revealed that in the Southern Ocean there is an intricate web of interleaving fronts that are not necessarily circumpolar in extent \citep{MooreEtAl1999,Hughes&Ash2001,KostianoyEtAl2003,Burl&Reason2006,DongEtAl2006,BillanyEtAl2010,GrahamEtAl2012}. However, oceanic eddies can induce SSH and SST gradients of similar or even greater magnitude than those associated with fronts. As differentiation (required to compute the gradient) tends to amplify ``noise", locally distinguishing between eddies and fronts is a difficult task and some kind of filtering or averaging should be employed.

The WHOSE method reduces the influence of eddies and other ``noise" using a denoising filter that specifically searches for ``step" like signals, that are the manifestation of oceanic fronts in SSH or SST, by exploiting the fact that step signals have non-zero kurtosis:
\begin{equation} \label{Eqn1.Kurtosis}
K(x) = \frac{\mathbb{E}\left [ \left (x-\overline{x}\right)^{4} \right] } { \left( \mathbb{E} \left [ \left ( x-\overline{x}\right)^{2}\right]\right)^{2} }-3,
\end{equation}   
where $x$ is a real random variable, $\mathbb{E}$ denotes the expectation operator and the overbar the mean \citep{Ravier&Amblard1998}. If the  ``noise" introduced by differentiation is assumed to have a normal-distribution (hence zero kurtosis) and the step-like ``signal" are non-normal (non-zero kurtosis) then a kurtosis criteria should be able to separate the signal from the noise. 

The algorithm has four primary steps:
\begin{enumerate}
\item [\textbf{Step 1}:]perform a wavelet decomposition in the \textit{space} domain in order to decompose the ADT into ``scales";
\item [\textbf{Step 2}:] at each wavelet scale, we determine if those coefficients are normal distributed -- and hence ``noise" or non-normal, and therefore signal, by determining if the kurtosis of the wavelet coefficients falls within bounds given by the Bienaym\'{e}-Chebyshev inequality (see \citet[p.4323]{Chapman2014}, Eqn. (4)). If the wavelet coefficients at that scale are found to be ``noise", then they are set to zero. Otherwise, they are retained without modification;
\item [\textbf{Step 3}:] For the remaining wavelet coefficients, apply a simple wavelet denoising procedure of \citet{Donoho&Johnstone1994}. Wavelet coefficients are set to zero where they are less than a threshold, $\lambda$, given by:
\begin{equation}
\lambda^{s} = \textrm{median} \left [ |d^{s}(x,t)|\right] \sqrt{2 \textrm{log}_{10}N}/0.6745
\end{equation}
where $s$ denotes the wavelet scale, $d$ the wavelet detail coefficients at scale $s$ and $N$ the number of points on the transect. 
\item [\textbf{Step 4}:] The inverse wavelet transform is used to reconstruct the SSH transect and gradient threshold is applied to determine the front locations. Following \cite{GrahamEtAl2012}, the gradient threshold is set to 0.1m/100km. 
\end{enumerate}  

This algorithm has been implemented in the open source Python language and the code is freely available from the author's Github repository (see Appendix A).


\section{Time Mean Frontal Behavior} \label{Section:Mean_Front_Positions}

\subsection{Frontal Occurrence Frequency and Time Mean Positions} \label{SubSection:Fitting_Heat_Maps}
\begin{figure*}[t]
   \centering  
  \includegraphics[width=40pc,height=20pc,angle=0]{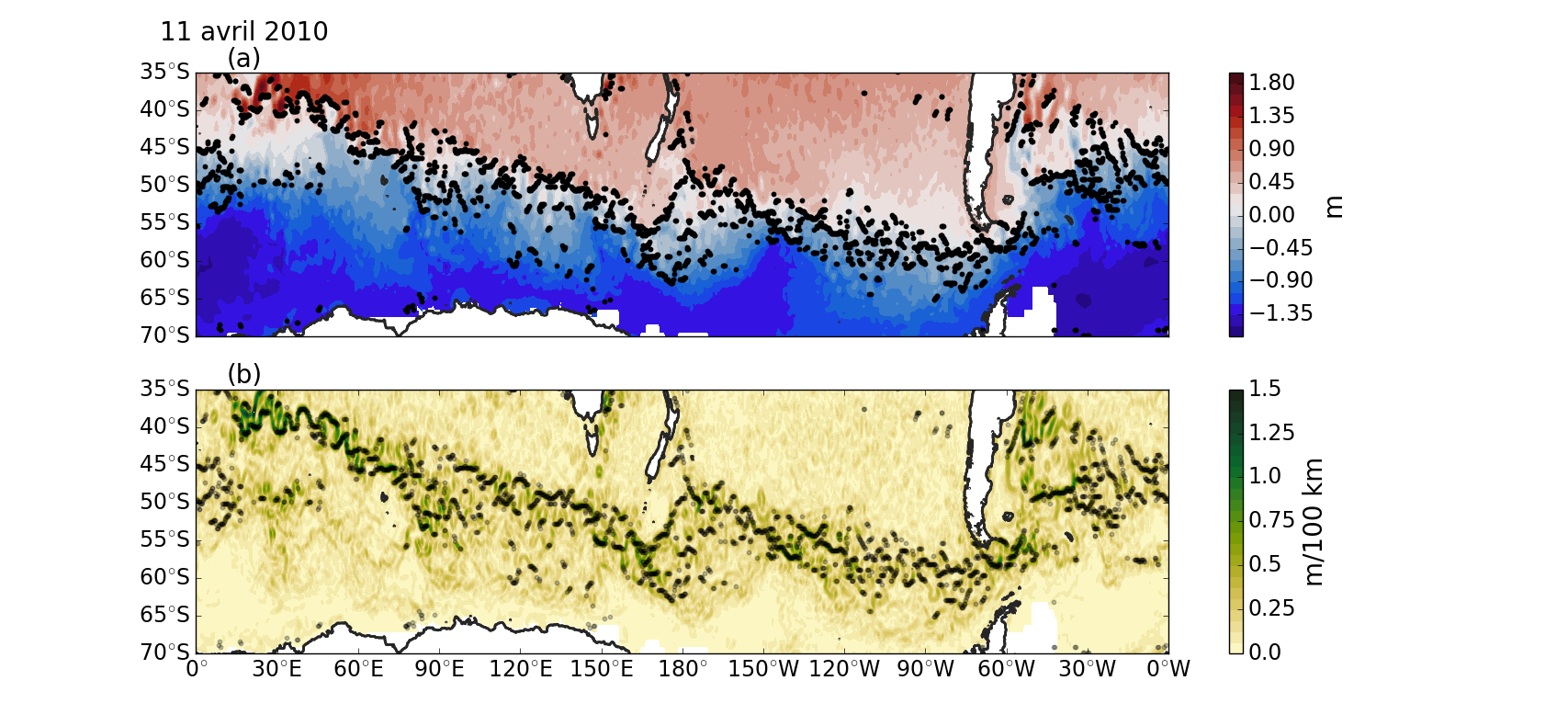}\\
  \caption{Snapshot of the (a) absolute dynamic topography (ADT); and (b) ADT gradient, on the 11 of April, 2010, overlaid with the locations of the detected fronts (closed circles: $\bullet$). Note that detected fronts outside of the Southern Ocean latitudes have been removed from the analysis (see text for details). }\label{Fig2:Snapshot_Detection}
\end{figure*}

The WHOSE method is applied to each daily ADT maps between 70$^{\circ}$S and 30$^{\circ}$S, producing 8035 daily maps of frontal location. An example is shown in Fig. \ref{Fig2:Snapshot_Detection}, along with a snapshot of the ADT (Fig. \ref{Fig2:Snapshot_Detection}(a)) and $|\nabla ADT|$ (Fig. \ref{Fig2:Snapshot_Detection}(b)). We see that frontal locations (circles), follow  large, yet zonally coherent ADT gradients. These daily maps of frontal location have been made freely available for download (see Appendix B). 

The detected fronts show significant non-zonal orientation, primarily due to steering of the flow by subsurface topography (e.g. between 150$^{\circ}$E and 180$^{\circ}$ at the Campbell Plateau) or high frequency variability induced by Rossby waves moving along the fronts, as described by \citet{Hughes1996}, mostly clearly seen in the Agulhas region, south of Africa (20$^{\circ}$E and 50$^{\circ}$E). Animations of frontal location indicate substantial high-frequency variability, including the aforementioned Rossby wave propagation, as well as the drift, splitting and merging identified in the idealized experiments conducted by \cite{Thompson2010}.

As implied by Fig. \ref{Fig2:Snapshot_Detection}, the WHOSE method produces frontal location maps that are essentially a three-dimensional (latitude, longitude and time) array of bits, where a ``TRUE" value indicates the presence of a front at that particular time and location. In contrast to contour methods, fronts defined in this way may not be present at all longitudes and at all times. The number of fronts present at any particular longitude ranges from as few as two (e,g, between longitudes 120$^{\circ}$E and 150$^{\circ}$E, south of Tasmania) to as many as 15 (e.g. between about 65$^{\circ}$W and 30$^{\circ}$W, downstream of Drake Passage). While perhaps more realistic, the fronts defined here are not as easy to analyze as those defined by contour type methods.     

To investigate the time-mean behavior of the fronts we form maps of frontal occurrence frequency (a kind of 2D histogram, or ``heat--map") by counting the number of times a front is found at each point in the domain over a certain time period. A sharp, localized peak in the heat--map indicates the presence of a persistent front with little time variability, while a more broad distribution of points around a peak indicates a meandering front. \cite{Chapman2014} has demonstrated that these heat--maps have a different qualitative character depending on the method used to detect the fronts. Contour methods, for example, show strongly meandering fronts down stream of topographic features, while gradient detection methods show a splitting of the fronts and a reorganization of the frontal structure.   

Before discussing the complicated flow structure in the Southern Ocean, we use a simple toy-problem to illustrate some aspects of the frontal frequency of occurrence heat-maps. Consider a meandering front with a meridional position, $y$, decomposed into a ``mean" component $ \overline{y}$ (that is possibly spatially variable), and a ``meandering" $y^{\prime}$ component:
\begin{equation} \label{Eqn:Stochastic_Front}
y(x;t) = \overline{y}\left( x;\right) + y^{\prime} (x;t)
\end{equation}  
where $x$ is the zonal coordinate and $t$ is time. Here $\overline{y^{\prime}}=0$, with the overline indicating the mean value. The meandering component $y^{\prime}$ is modeled as a spatially auto-correlated AR(1) stochastic process:
\begin{equation} \label{Eqn:Front_Meander}
y^{\prime} (x_{i+1};t) = \rho y^{\prime} (x_i,;) + \epsilon
\end{equation}  
where $\rho$=0.9 is the auto-correlation parameter and $\epsilon \sim \mathcal{N}\left(0,\sigma^{2}\right)$ is uncorrelated white-noise drawn from a normal distribution with zero mean and a standard deviation $\sigma=\sigma(x)$ that can vary spatially. 

\begin{figure*}[t]
   \centering  
  \includegraphics[width=40pc,height=20pc,angle=0]{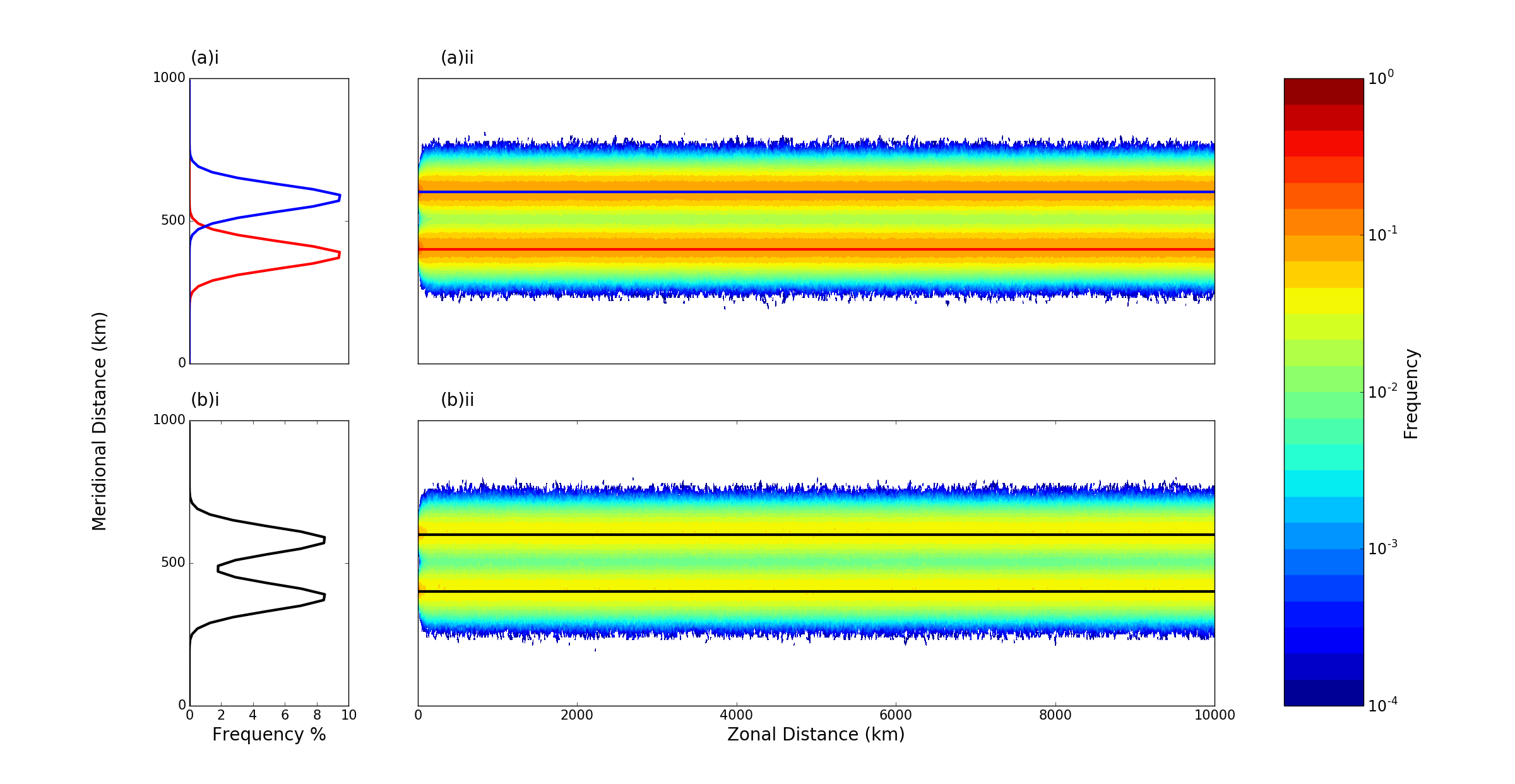}\\
  \caption{Multiple fronts versus a single bi-stable front in a toy model. (a)i The meridional position PDFs for two synthetic fronts centered at $y$=400km and 600km; (a)ii the (normalised) frontal frequency of occurance heat-map for the situation with two stochastically meandering fronts. The blue (red) solid lines indicate the mean frontal positions;  (b)i the meridional position PDFs for a single synthetic front with two stable positions ($y$=400km and 600km); and (b)ii the (normalised) frontal frequency of occurance heat-maps for the single bi-stable front situation. The black solid lines indicate the two mean positions}\label{Fig3b:Flipping_vs_Multiple_Fronts}
\end{figure*}

Fig. \ref{Fig3b:Flipping_vs_Multiple_Fronts} shows the heat-maps generated from 10,000 realizations of the toy-problem for two different situations: two fronts the meander about constant mean meridional positions $\overline{y}_1 =$ 400km, and $\overline{y}_2=$ 600km (panels (a)i--ii); and a single meandering bi-stable front with two preferred mean locations at 400km and 600km. In the latter case the front randomly flips between the two mean positions, as in \cite{Chapman&Hogg2013}. In both cases, the standard deviation, $\sigma$, of the meandering component is a constant 25km. We see in Figs. \ref{Fig3b:Flipping_vs_Multiple_Fronts}a and b that the heat maps produced are virtually indistinguishable, despite the different underlying frontal structures that have been used to produce them. As such, we note that our heat-map method is unable to distinguish between an ephemeral front that changes its location or two stable, neighboring fronts. However, as noted by \cite{Chapman&Morrow2014a}, the fronts that exhibit bi-stable behavior are somewhat rare in the Southern Ocean outside of particular regions. As such, herein we interpret two adjacent peaks in a heat-map as indicating two different fronts and not a single ephemeral front.

\begin{figure*}[t]
   \centering  
  \includegraphics[width=40pc,height=20pc,angle=0]{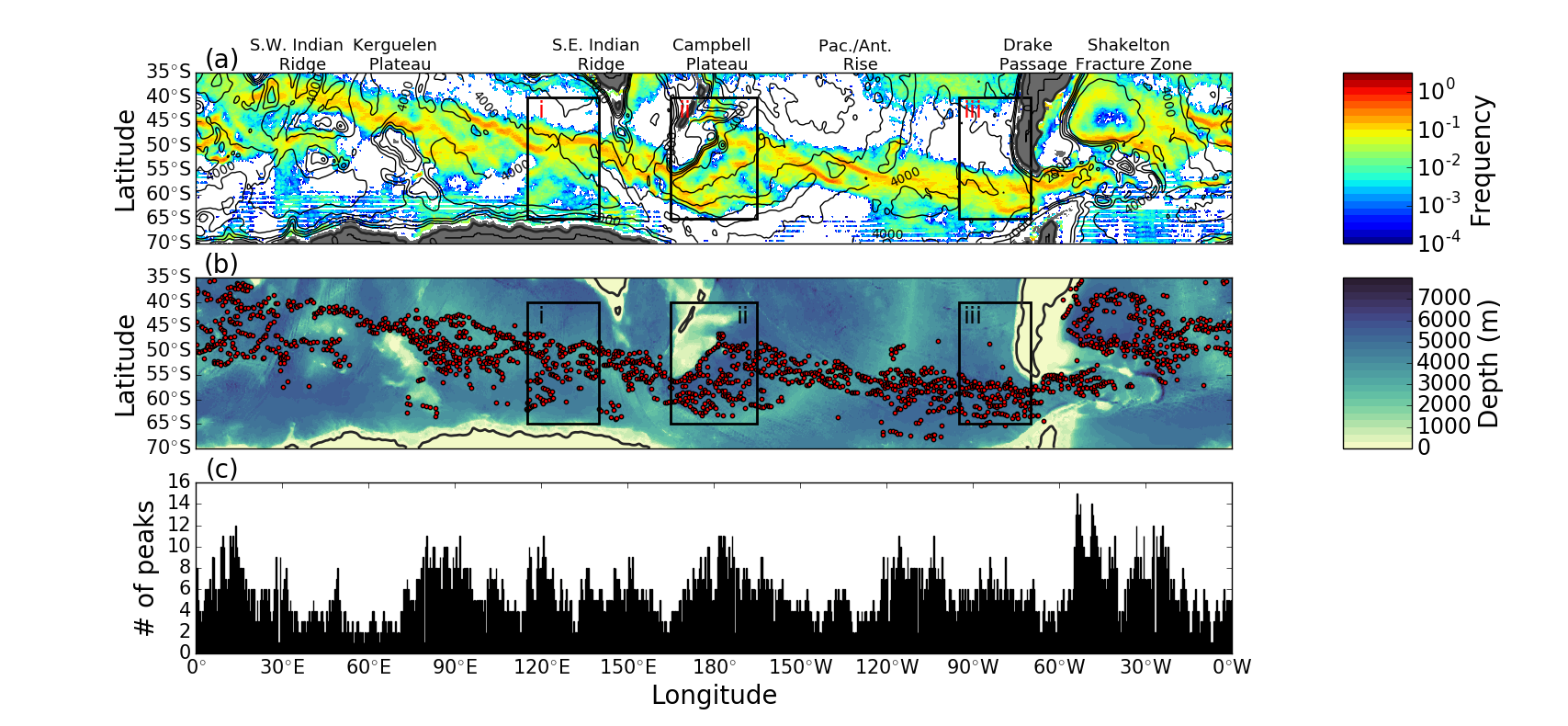}\\
  \caption{The time-mean structure of the Southern Ocean fronts. (a) The (normalised) frontal frequency of occurance ``heat maps" for the period 1993--2015 (note the logarithmic color scale); (b) the time-mean frontal locations determined using the curve fitting proceedure (red $\bullet$), overlaying the ocean bathemetry from the ETOPO01 dataset (shaded contours); and (c) the time--mean number of fronts at each longitude. The boxes in panels (a) and (b) indicate the zoomed regions of Fig. \ref{Fig4:Histograms_Zoomed}}\label{Fig3:Histograms_Centres_Njets}
\end{figure*}

The heat-map obtained from the WHOSE method for the period 1993--2015 is shown in Fig. \ref{Fig3:Histograms_Centres_Njets}a, which reveals a complicated frontal structure that changes throughout the Southern Ocean. In certain regions, for example south of Australia (between about 90$^{\circ}$E and 130$^{\circ}$E -- box \textit{i}), the heat maps indicate one or two strongly persistent fronts. Other regions, for example near the Campbell Plateau south of New Zealand (170$^{\circ}$E to 170$^{\circ}$W -- box \textit{ii}) show multiple persistent fronts with clear peaks in the heat--map, but with a broad distribution of points about those peaks, indicative of meandering fronts. Finally, there are regions, such as the eastern Pacific (90$^{\circ}$W to 70$^{\circ}$E -- box \textit{iii}) where the frontal structure consists of a broad distribution of points with no dominant peak, indicative of either strongly meandering fronts or numerous, closely packed fronts. The latter region corresponds to the region identified by \cite{ThompsonEtAl2010} where the interior potential vorticity structure is homogenized near the surface and the PV gradients are weak.

Although heat-maps provide a qualitatively attractive framework for understanding the ACC's frontal system, making robust \textit{quantitative} arguments based on these maps is more difficult. To tackle this problem, we fit to each meridional transect of the heat-map an arbitrary number of skew normal functions \citep{Azzalini&Capitanio1999}:
\begin{equation}
f(\phi;A,\overline{\phi}_j,\sigma,\gamma) = Ae^{-\frac{(\phi-\overline{\phi})^{2}	}{2\sigma^{2}}} \left \{1 + \textrm{erf} \left [ \frac{\gamma \left(\phi-\overline{\phi}_j \right) }{\sqrt{2}\sigma} \right ] \right\},
\end{equation}
where $\textrm{erf}$ is the error function, $A$ the amplitude, $\phi_j$ the mean latitude, $\sigma$ the standard deviation and $\gamma$ the ``skew" parameter. Skew normal curves differ from a standard normal curve frontal by the presence of asymmetry, measured by the skew parameter $\gamma$. A positive (negative) $\gamma$ results in an elongated tail to the right (left) of the mean (here to the north). Thus, at each longitude, $\lambda$, in the domain, the frontal occurrence heat-map is approximated as:
\begin{equation}
P(\lambda;\phi) = \sum^{n}_{i=0} f(\phi;A_i,\overline{\phi}_i,\sigma_i,\gamma_i).
\end{equation}  
Each component function $f_i$ shall be interpreted as a separate front at that longitude. To determine the number of fronts $n$, we use the Akaike Information Criterion (AIC) \citep{Akaike1974,BurnhamEtAl2011}, which for a heat--map transect modelled by the sum of $m$ skewed-normal functions, is:
\begin{equation} \label{Eqn:AIC}
\textrm{AIC}_m = 8m + \log{\mathcal{L}} 
\end{equation}  
where $\mathcal{L}$ is the maxima of the \textit{liklihood} function (a measure of the goodness-of-fit of the model to the data). The AIC quantifies the trade-off between how well a model fits the data and the number of parameters in the model. A model with a lower AIC is preferred. The 8$m$ term in Eqn. \ref{Eqn:AIC} effectively penalises fitting additional functions (hence another frontal peak). Starting with $n$=1, additional skew-normal components are added until the AIC reaches a local minima. 

Using the skew-normal fits we are able to determine, at every the longitude in the domain, the number of fronts present, $n$, their mean latitude, $\overline{\phi}$ and their meandering standard deviation $\sigma$. The time-mean positions of the fronts are shown in Fig. \ref{Fig3:Histograms_Centres_Njets}b, and are in agreement with previous studies \citep[eg.]{GrahamEtAl2012}. The number of distinct peaks detected is shown in Fig. \ref{Fig3:Histograms_Centres_Njets}c, where it is clear that the number of fronts is highly variable, with as many as 15 peaks, to as few as a one. ``Splitting" behaviour (where the number of fronts \textit{increases} downstream) occurs primarily downstream of large topographic features such as the Kerguelen Plateau (labelled in Fig. \ref{Fig3:Histograms_Centres_Njets}), while the inverse phenomena of ``merging" (the number of fronts \textit{decreases} downstream) occurs primarily in regions where the flow is steered or constricted by topography. Note that at approximately 50$^{\circ}$W, the number of fronts reaches its global maxima due, in part, to the splitting and northward deviation of the SAF along the coast of South America, previously identified by \cite{Sokolov&Rintoul2009a}.

\begin{figure}[t]
   \centering  
  \includegraphics[width=18pc,height=18pc,angle=0]{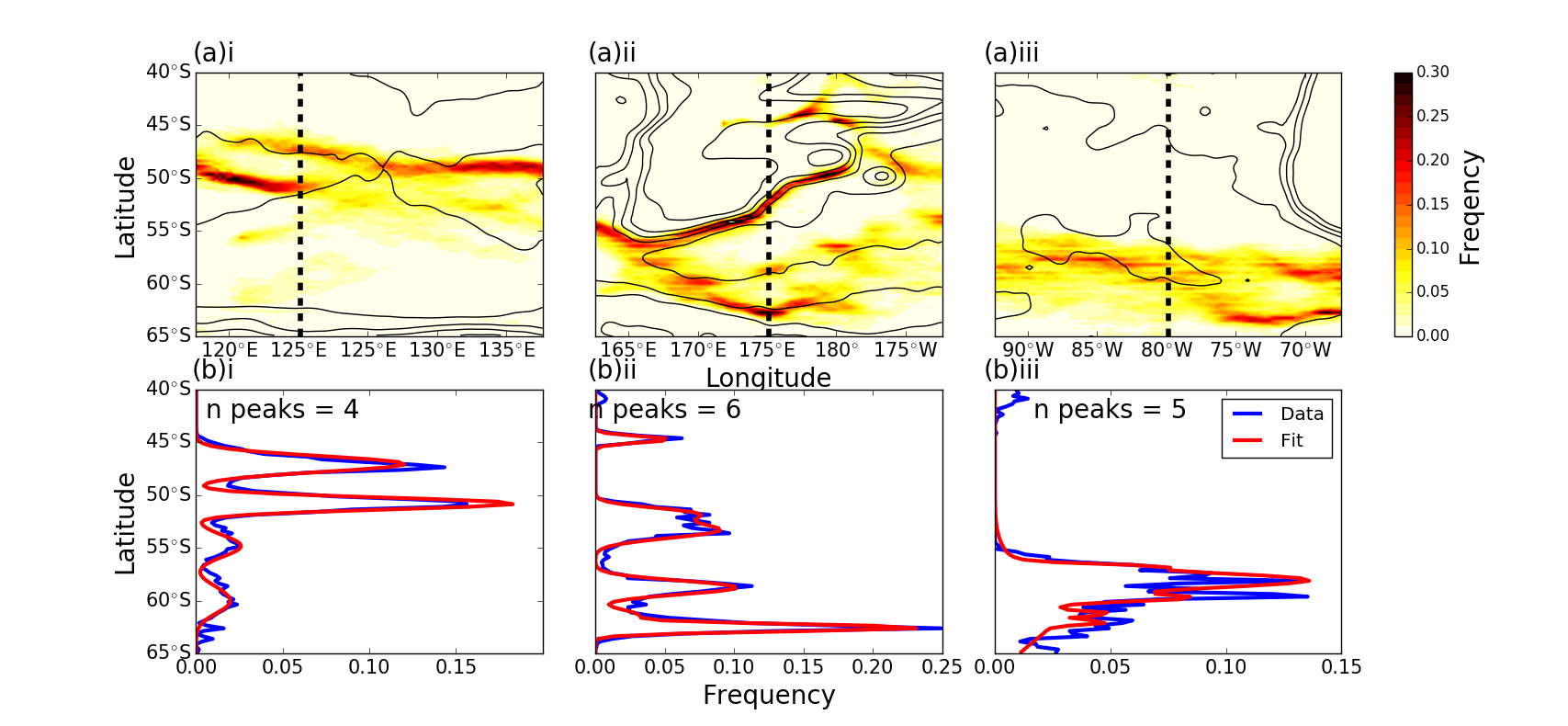}\\
  \caption{Local variations in the frontal structure. (a)i---iii Zoomed view of the frontal frequency of occurance heat-maps within the labelled boxes of Fig. \ref{Fig3:Histograms_Centres_Njets} (note that the linear colorscale) with the bathymetric contours (black solid lines: C.I. 1000m); and (b)i---iii meridional transects of the frontal frequency of occurance along the transects indicated by the dotted lines in upper panels (blue) and the associated skew-normal curve fit (red).}\label{Fig4:Histograms_Zoomed}
\end{figure}

While the method of fitting simple functions to the heat-maps has allowed us to estimate the number of fronts at each longitude, there remains ambiguity about the exact nature of the underlying frontal structure. Illustrating this fact, Fig. \ref{Fig4:Histograms_Zoomed}a shows zoomed views heat-maps in the boxes indicated in Fig. \ref{Fig3:Histograms_Centres_Njets}, while Fig. \ref{Fig4:Histograms_Zoomed}b shows the transects of the heat-maps and their associated skew-normal fits along the dashed--lines in Fig \ref{Fig4:Histograms_Zoomed}a. These regions have be chosen to demonstrate how the frontal structure can vary throughout the Southern Ocean. 
     
Box \textit{i}, a region near the zonally oriented Southeast Indian Ridge, shows a situation with two strong, persistent fronts (at approximately 47$^{\circ}$S and 52$^{\circ}$S) and two less persistent fronts to the south (approximately 55$^{\circ}$S and 60$^{\circ}$S), each with clearly identifiable, isolated maxima in the frontal occurrence transect. Along the chosen transect, it is clear that the fitting method identifies the four fronts and is able to reproduce the amplitudes and standard deviations (that is, the persistence and meandering of fronts). 

The region in box \textit{ii}, south of New Zealand and the Campbell Plateau, has more complex bathymetry than that of box \textit{i} and the frontal structure is itself more complicated. In this region, the fitting procedure identifies six fronts. However, on closer inspection, the southern region of high frontal frequency (centered near $\sim$64$^{\circ}$S) as well as the front that follows the edge of the Campbell Plateau (at $\sim$53$^{\circ}$S) are each represented by two overlapping skew--normal functions, one that captures the dominant peak and broad shape of the that cluster and the another that produces a secondary peak in frontal frequency ($\sim$52$^{\circ}$S and $\sim$57$^{\circ}$S). 

The frontal structure in box \textit{iii}, in the eastern South Pacific, a region of generally flat bottom topography, differs from that of boxes \textit{i} and \textit{ii}. It is illustrative to compare the heat-map in this region, which shows a broad area of elevated frontal occurrence frequency, with the snapshots in Fig. \ref{Fig2:Snapshot_Detection}, where the numerous braided fronts are evident. Further inspection reveals that the fronts here are ephemeral and closely packed together. The lack of persistence of these fronts and their close proximity acts to smear out the heat-maps over a swath of latitudes. The transect through this region (Fig. \ref{Fig4:Histograms_Zoomed}b(iii)), indicates a region of high frontal recurrence spread over a range of latitudes, with two dominant peaks separated by less than 2$^{\circ}$ of latitude, as well as numerous sub-dominant peaks. The heat-map is noisy in this region and the fitting method, while able to effectively identify the locations of the two dominant peaks, it does not capture all the sub-dominant peaks, instead representing them with a large enough standard deviation to encompass their regions of enhanced frontal occurrence frequency. 

\subsection{Connection with fronts defined by contours of sea-surface height} 

\begin{figure*}[t]
   \centering  
  \includegraphics[width=40pc,height=20pc,angle=0]{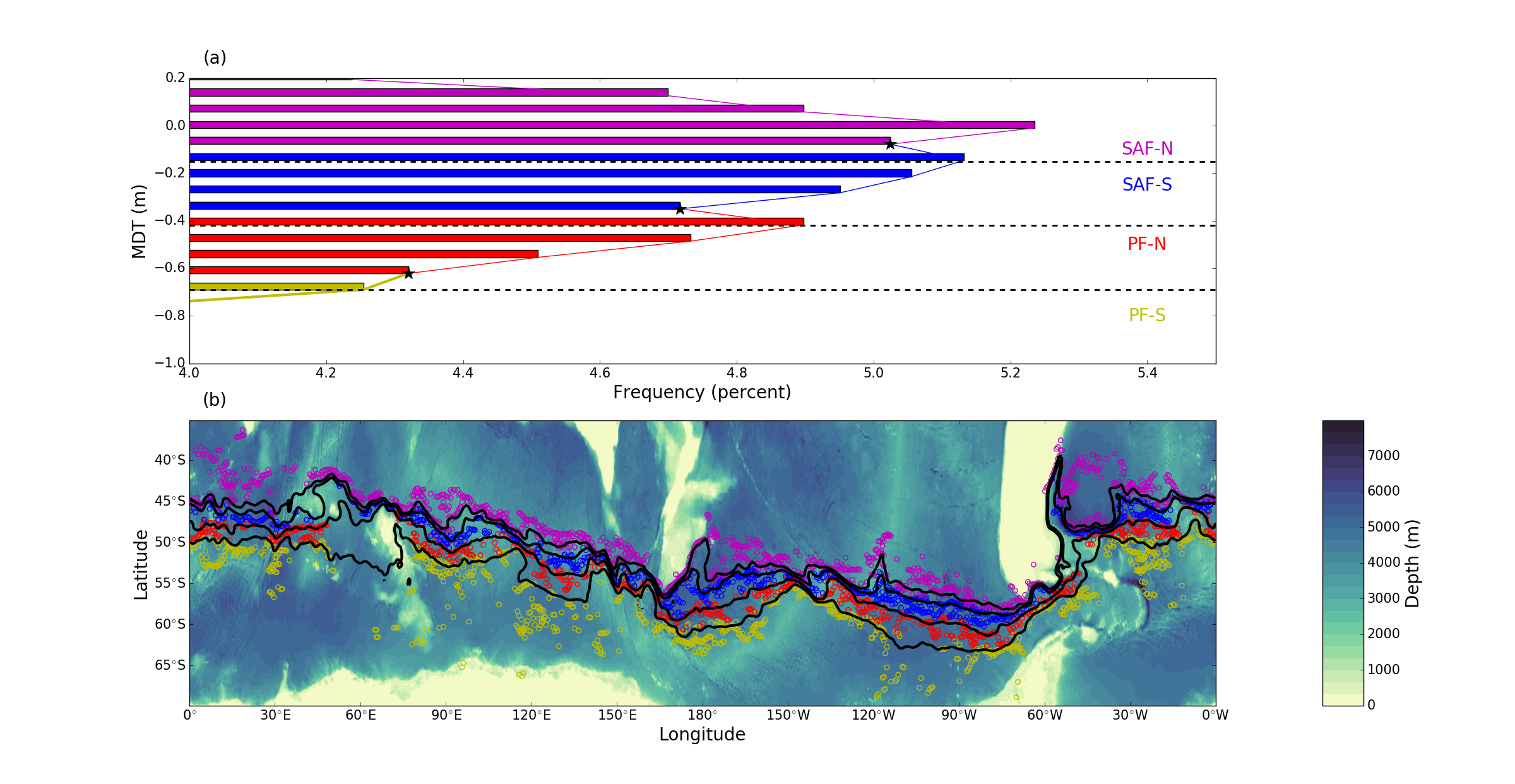}\\
  \caption{The relationship of the time-mean frontal positions and the MDT contours (a) the histogram of the MDT found at each time-mean front location (bin width of 0.07 dyn m). The coloring of the curve indicates the region in MDT-space associated with each named front, while the $\star$ indicate the change from one frontal regime to another; and (b) the time-mean locations of the fronts from Fig. \ref{Fig3:Histograms_Centres_Njets}b (colored open circles) with the position of those same fronts defined using contours of MDT (solid lines) overlying the ocean bathymetry. The fronts are color coded from north to south as:  SAF-N: magenta, SAF-S: blue; PF-N: red; and PF-S: yellow.}\label{Fig5:Front_Locs_With_MDT}
\end{figure*}

We now compare the fronts detected by the WHOSE method with those defined by altimetrically derived dynamic height contours \citep{Sokolov&Rintoul2007}. To proceed, we determine the value of the MDT from the \cite{RioEtAl2014} dataset at each of the time-mean frontal locations in Fig. \ref{Fig3:Histograms_Centres_Njets}b (a similar analysis using the time-varying ADT for the time varying frontal positions is presented in section \ref{Section:Variability_In_Front_Positions}). The normalized histogram of the MDT at the mean frontal locations, computed with an MDT bin spacing of 0.07dyn m, is shown in Fig. \ref{Fig5:Front_Locs_With_MDT}a. Peaks in the histogram (dashed lines in Fig. \ref{Fig5:Front_Locs_With_MDT}a) are used to define a MDT value with which we can identify a front.


Four distinct peaks in the MDT histogram are identified. Based on comparisons with \cite{Sokolov&Rintoul2007}, we label the detected fronts the Subantarctic Front (SAF) and the Polar Front (PF), both of which have a northern and a southern branch (denoted SAF-N \& SAF-S in the case of the Subantarctic Fronts and PF-N and PF-S for the Polar Fronts). The values of the MDT used to define the fronts are listed in Table \ref{Table1:ADTValues}, along with comparable values from other studies\footnote{Note that there is no clearly defined definition of the SAF and the PF and hence the naming convention differs between studies. For example \cite{Volkov&Zlotnicki2012} and \cite{LanglaisEtAl2011} define the SAF-S at -0.4 dyn m, where we have defined the PF-N  and \cite{LanglaisEtAl2011} defines the PF-S at -0.68 dyn m which here that MDT value is associated with the PF-N}. Note that the sAACf is not analyzed in this study. 

\begin{table*}[t]

\caption{ADT Values used to define the fronts in this and two others studies that make use of comparable data.}  \label{Table1:ADTValues}
\begin{center} 
\begin{tabular}{ccccc}
\hline\hline
Front &  ADT (dyn m) & ADT (dyn m)                     & ADT (dyn m)                 &  ADT (dyn m)   \\
      &  This study  & Langlais \textit{et al.} (2011) & Volkov \& Zlotnicki (2012)  & Kim \& Orsi (2014) \\

\hline \hline
 SAF-N      &   -0.01   &  N/A    &   N/A  &  -0.03 \\
 SAF-S      &   -0.15   &  -0.16  &   -0.2  &  N/A  \\
\hline
 PF-N       &   -0.42   &  -0.40  &   -0.4  &   N/A  \\
 PF-S       &   -0.69   &  -0.68  &   -1.0  & -0.61  \\
\hline \hline
\end{tabular}
\end{center}
\end{table*}

It is clear in Fig. \ref{Fig5:Front_Locs_With_MDT}b that not all of the detected frontal positions lie on one of the three MDT contours. In order to associate each time-mean front location with one of the four fronts defined here, we note that each of the peaks in the histogram are separated by a local minima (black $\star$ in Fig. \ref{Fig5:Front_Locs_With_MDT}a). Following \cite{Azzalini&Torelli2007}, we use the minima of histogram to bound the time-mean front locations as belonging to one of the SAF-N, SAF-S, PF-N or PF-S. Consider the PF-N, centered in MDT space on the peak of the histogram at -0.40dyn m. This maxima is bounded by minima at MDTs -0.23 dyn m and -0.65 dyn m. All frontal locations falling in MDT space between these values are labeled as belonging to the PF-N. The time-mean frontal locations classified in this way are shown in Fig. \ref{Fig5:Front_Locs_With_MDT} as open circles, colored to indicate their classification.  

From a time-mean perspective, the selected MDT contours (solid lines in Fig. \ref{Fig5:Front_Locs_With_MDT}b) approximate well the time-mean location of the WHOSE fronts. This is particularly true for the SAF-S and PF-N: the bias in the frontal positions estimated by the MDT contour is less than 0.5$^{\circ}$ for both of these fronts. Although the bias is larger for the SAF-N and PF-S, the increased error occurs due to the presence of WHOSE fronts found far to the north or south that are not in the main core of the ACC. When these points are excluded, the bias for both SAF-N and PF-S approaches those of the SAF-S and PF-N.  

Despite the good results obtained by MDT contour fits, the qualitative frontal structure revealed by the contours differs somewhat from that revealed by the WHOSE fronts. Firstly, as shown in Fig. \ref{Fig3:Histograms_Centres_Njets}c, the number of fronts determined by the skew-normal fits vary substantially throughout the Southern Ocean. For example, note in Fig. \ref{Fig5:Front_Locs_With_MDT}b the sub-branch of one front (for example, the SAF-S) may itself split into sub-sub-branches. There are as many as five sub-branchs of the SAF-N, and as many as four SAF-S and PF-N sub-branches. It is also notable that differently labelled fronts do not split or merge in the same locations, nor do splitting or merging events seem to be consistently controlled by topography. Downstream of the Campbell Plateau (at 175$^{\circ}$W-- see Fig. \ref{Fig4:Histograms_Zoomed}b--ii) an area where the fronts are strongly steered by topography,  SAF-N undergoes a clear splitting events while the SAF-S does not. In other locations, such as upstream of the Kergeulen Plateau,  the MDT contours imply the presence of fronts where none have been detected (Fig. \ref{Fig5:Front_Locs_With_MDT}b). \cite{Thompson2010} has shown in a series of idealized numerical experiments that splitting and merging behavior depends strongly upon topographic details, such as amplitude, orientation and length scale of the underlying topographic feature, and similar dynamics may be in play in here. 

Secondly, the time-mean frontal locations determined here do not always follow the MDT contours chosen to represent them. The local misfit between the two different definitions is greater for the SAF-N and PF-S than for the other frontal branches due to the fact that both these fronts encompass a wider range of dynamic heights and latitudes. However, similar misfits can be found in both the SAF-S and PF-N, albeit less frequently. We will return to this point in section \ref{Section:Variability_In_Front_Positions}.

\subsection{Spatial variability of the frontal structure}

The Southern Ocean is zonally asymmetric. Mesoscale activity, the primary orientation of the mean flow and, as is clear in Fig. \ref{Fig3:Histograms_Centres_Njets}, the structure of the fronts, all vary with longitude \citep{Rintoul&Garabato2013}. Bathymetry, which both steers the ACC currents and acts to generate regions of elevated turbulent activity, is thought to influence fronts in two ways: firstly, fronts which pass over regions of with large gradients in bottom topography (and hence, modified background potential vorticity gradients) are constrained in their movement and their capacity to meandering about their mean locations is greatly reduced \citep{SokolovAndRintoul2009II}; secondly, since mesoscale turbulence is generated preferentially in regions downstream of large bathymetric features \citep{WilliamsEtAl2007,ChapmanEtAl2015} it has also been suggested that fronts may show increased variability in these regions \citep{LanglaisEtAl2011,Chapman2014}. However, there is some disagreement about how this variability manifests: \cite{LanglaisEtAl2011} suggests coherent meandering of the frontal filaments, while \cite{Chapman2014} provides evidence fronts splitting into additional smaller scale sub-fronts in these regions.

\begin{figure}[t!]
   \centering  
  \includegraphics[width=18pc,height=18pc,angle=0]{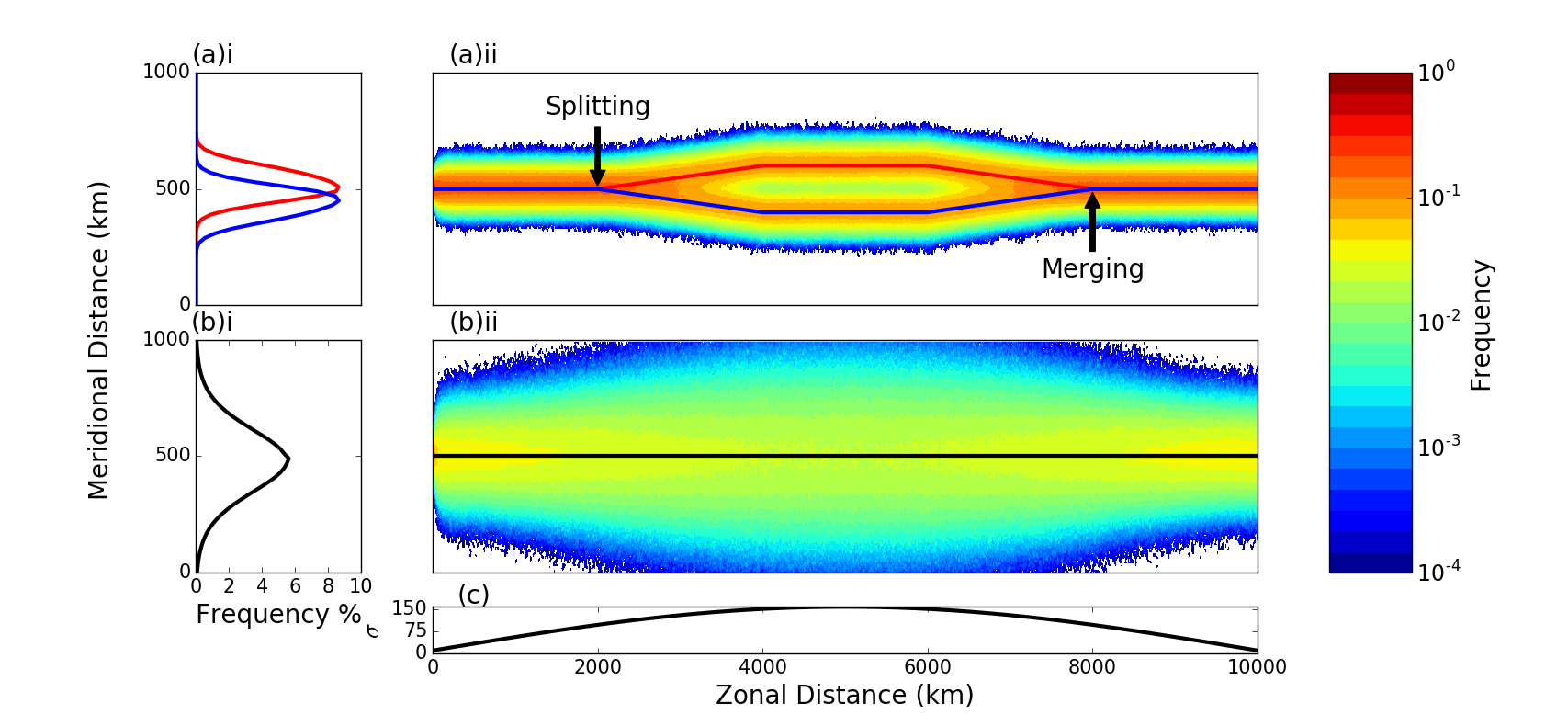}\\
  \caption{Splitting/merging compared with variable meandering in a toy-model: (a)i The meridional position PDFs for two synthetic fronts that undergo splitting and merging. The red (blue) line indicates the front that splits to the north (south); (a)ii the (normalised) frontal frequency of occurance ``heat maps" for synthetic fronts that split and merge. Solid red (blue) line indicates the mean front position; (b)i as in (a)i for a single front with a increasing than decreasing meander amplitude; (b)ii as in (a)ii for the situtation with a single front with increasing then decreasing meandering amplitude; (c) the spatial variation of the standard deviation of the front  in panels (b)i and (b)ii.   }\label{Fig8:Splitting_Merging_vs_Meandering}
\end{figure}

To illustrate how these differing kinds of variability manifest in our heat-maps, we turn again to the toy-problem presented in Sec. \ref{Section:Mean_Front_Positions}. In Fig. \ref{Fig8:Splitting_Merging_vs_Meandering} frontal frequency heat-maps for two different situations are represented: a ``splitting/merging" case (Fig. \ref{Fig8:Splitting_Merging_vs_Meandering}a) where a single front with a constant meandering amplitude of $\sigma$=25km, splits at a particular location into two distinct fronts (indicated by the solid lines in Fig. \ref{Fig8:Splitting_Merging_vs_Meandering}a(ii)) before merging again, ; and a ``variable meandering" case  (Fig. \ref{Fig8:Splitting_Merging_vs_Meandering}b): a single front with a spatially varying meander amplitude that has the form:
\begin{displaymath}
\sigma(x) = A_{v}\cos \left( \frac{\pi}{L_x}x \right) + A_0
\end{displaymath}
where $A_0$=10km is the constant background meandering amplitude and $A_v$=150km is the spatially varying meander amplitude (shown in Fig. \ref{Fig8:Splitting_Merging_vs_Meandering}c).

Unlike the situation described in Sec. \ref{Section:Mean_Front_Positions} where the manifestation of the different physical phenomena were indistinguishable, the heat-maps shown in Fig. \ref{Fig8:Splitting_Merging_vs_Meandering} have a different structure, although there are similarities. For the splitting/merging case, the heat-map broadens and reduces in magnitude after the splitting event. Additionally, after the splitting occurs, (indicated by the arrow in Fig. \ref{Fig8:Splitting_Merging_vs_Meandering}a(ii)) there is a transition from a single peak in the heat-map to two clearly defined peaks separated by a minima, with the reverse occurring where the fronts merge. Similarly, in the variable meandering case the heat-map broadens and reduces in amplitude as $\sigma$ increases. However, there is no development of an additional peak in the heat-map. As such, we can distinguish between splitting/merging and variable meandering  with a knowledge of two key-parameters, the meander amplitude, measured by $\sigma$, and the number of peaks in the heat-map at any particular longitude.   
 
\begin{figure}[b!]
   \centering  
  \includegraphics[width=18pc,height=18pc,angle=0]{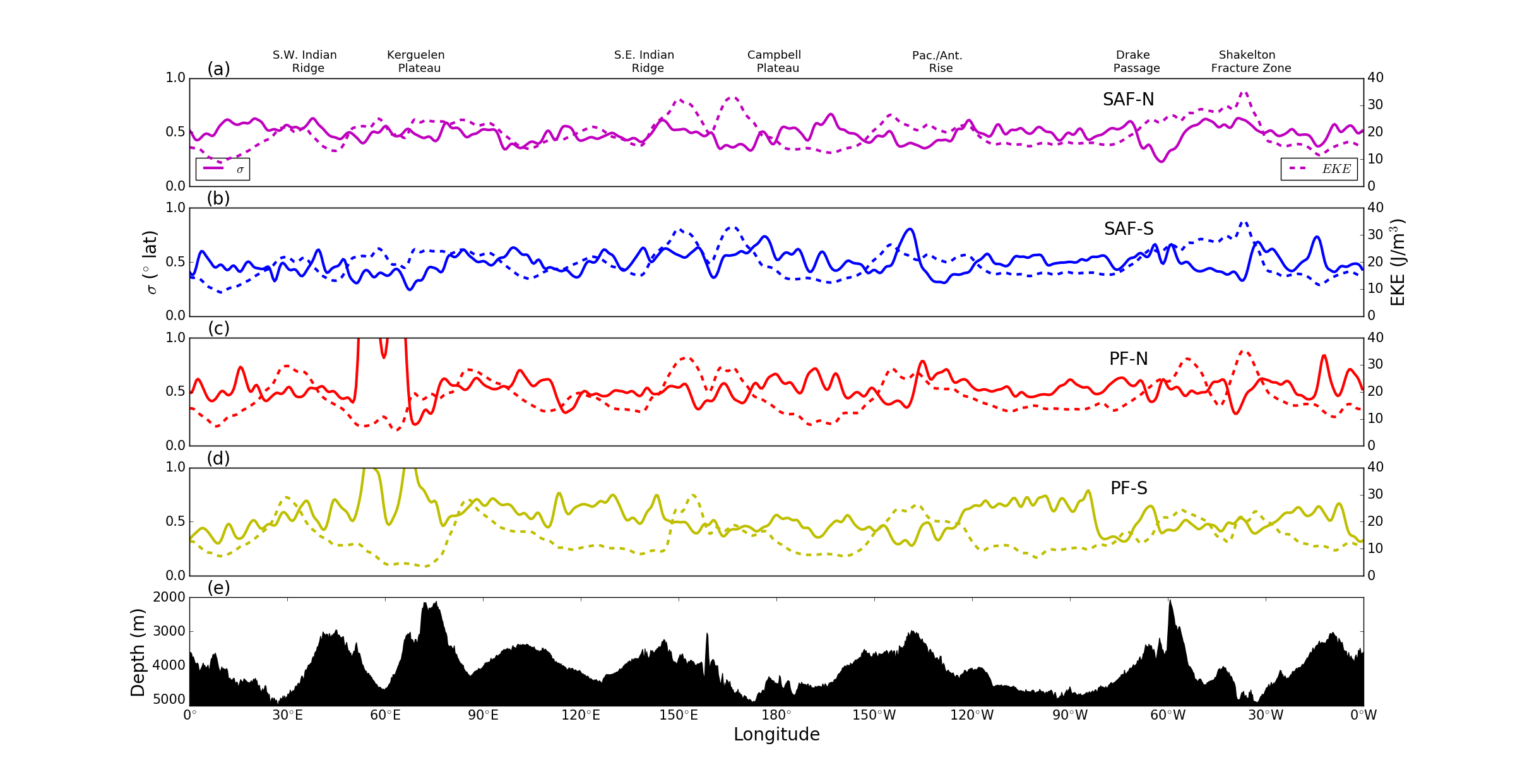}\\
  \caption{The average standard deviation $\sigma$ determined by the fitted skew-normal distributions to the frontal freqency of occurance maps (solid lines) and the eddy kinetic energy (dashed lines) for the (a) SAF-N; (b) SAF-S; (c) PF-N; and (d) PF-S. The bathymetry averaged across the ACC is shown in (e) for reference.}\label{Fig7:Std_dev_by_frontal_label}
\end{figure}

Recall that at each longitude, each fitted skew-normal function has an associated standard-deviation, $\sigma$. This standard deviation is plotted in Fig. \ref{Fig7:Std_dev_by_frontal_label} for each labeled front (averaged across all points that `belong' to a particular front), together with the averaged eddy kinetic energy (EKE), defined as:
\begin{equation}
EKE = \frac{1}{2} \left(\mathbf{u}^{\prime} \cdot \mathbf{u}^{\prime} \right)
\end{equation}
where $\mathbf{u}^{\prime} = \left(u^{\prime},v^{\prime} \right)$ is the horizontal surface perturbation velocity obtained from the AVISO SLA data. The quantities are determined at each of the time-mean frontal locations, determined from the altimetric surface velocity anomalies. 


All four fronts show a mean standard deviation of about 0.5$^{\circ}$ across all longitudes. Although there are variation from the mean value at various locations throughout the Southern Ocean, notably a large increase in the standard deviation upstream of the Kerguelen Plateau at around 60$^{\circ}$E where the ACC is influenced by the Aghulas current, and a moderate decrease in regions where the flow is strongly steered by bathymetry, such as through Drake Passage, at the Pacific Antarctic Rise and the Campbell Plateau. However, over the entire circumpolar circuit, there is little variation in the standard deviation. There is also no significant correlation between the EKE and standard deviation. In contrast, it is clear from Fig. \ref{Fig3:Histograms_Centres_Njets}c from that the number of fronts varies substantially throughout the Southern Ocean, and that these variations are spatially coherent.

The fact that the standard deviation remains roughly constant throughout Southern Ocean while the number of fronts varies substantially, provides some basis to claim that the dominant manifestation of spatial variability in the Southern Ocean is not a change in the capacity of fronts to meander, but is instead the splitting of the fronts into sub-fronts. These results are consistent with the analysis of the isopyncal potential vorticity (IPV)  in a high-resolution ocean model by \cite{ThompsonEtAl2010}, who show rearrangement of the IPV structure consistent with frontal splitting in the regions corresponding to boxes \textit{ii} and \textit{iii} in Fig. \ref{Fig4:Histograms_Zoomed}. 

We now seek to determine the long term changes in frontal location and ascertain if the frontal structure is sensitive to changes in atmospheric forcing. 

\section{Long term trends in Southern Ocean fronts positions} \label{Section:Variability_In_Front_Positions}

Have the location of the Southern Ocean fronts changed over the satellite era, and, if so, how are the changes related to variations in atmospheric forcing? The literature does not provide a clear answer to these questions. Certain studies, generally those that use a fixed SSH contour to define the location of the front, have found evidence of a southward shift in the frontal positions of between 0.5$^{\circ}$ and 1.0$^{\circ}$ \citep{SalleeEtAl2008,SokolovAndRintoul2009II,Kim&Orsi2014}, although the magnitude of that that shift is highly spatially variable. In contrast, studies employing local frontal definitions have found minimal long term temporal variability in the locations of the Southern Ocean fronts \citep{GrahamEtAl2012,ShaoEtAl2015,FreemanEtAl2016} although there is still the suggestion that fronts may shift with a limited geographical range, or may respond to modifications of the surface forcing that accompany changes in climate modes. 

In this section we employ our methodology to the problem of determining the long term trend in ACC frontal positions. In doing so, we will attempt to shed light on the discrepancy between those studies that find significant temporal shifts in frontal position, and those that find none. 

\subsection{Do fronts follow dynamic height contours?}

In section \ref{Section:Mean_Front_Positions} we showed that the time-mean front positions determined by our methodology are well approximated by the contours of MDT. However, previous work has questioned if the MDT contours are capable of tracking  \citep{GrahamEtAl2012}. Are contour methods the right choice for studying the temporal variability of Southern Ocean fronts? 

\begin{figure*}[t]
   \centering  
  \includegraphics[width=40pc,height=20pc,angle=0]{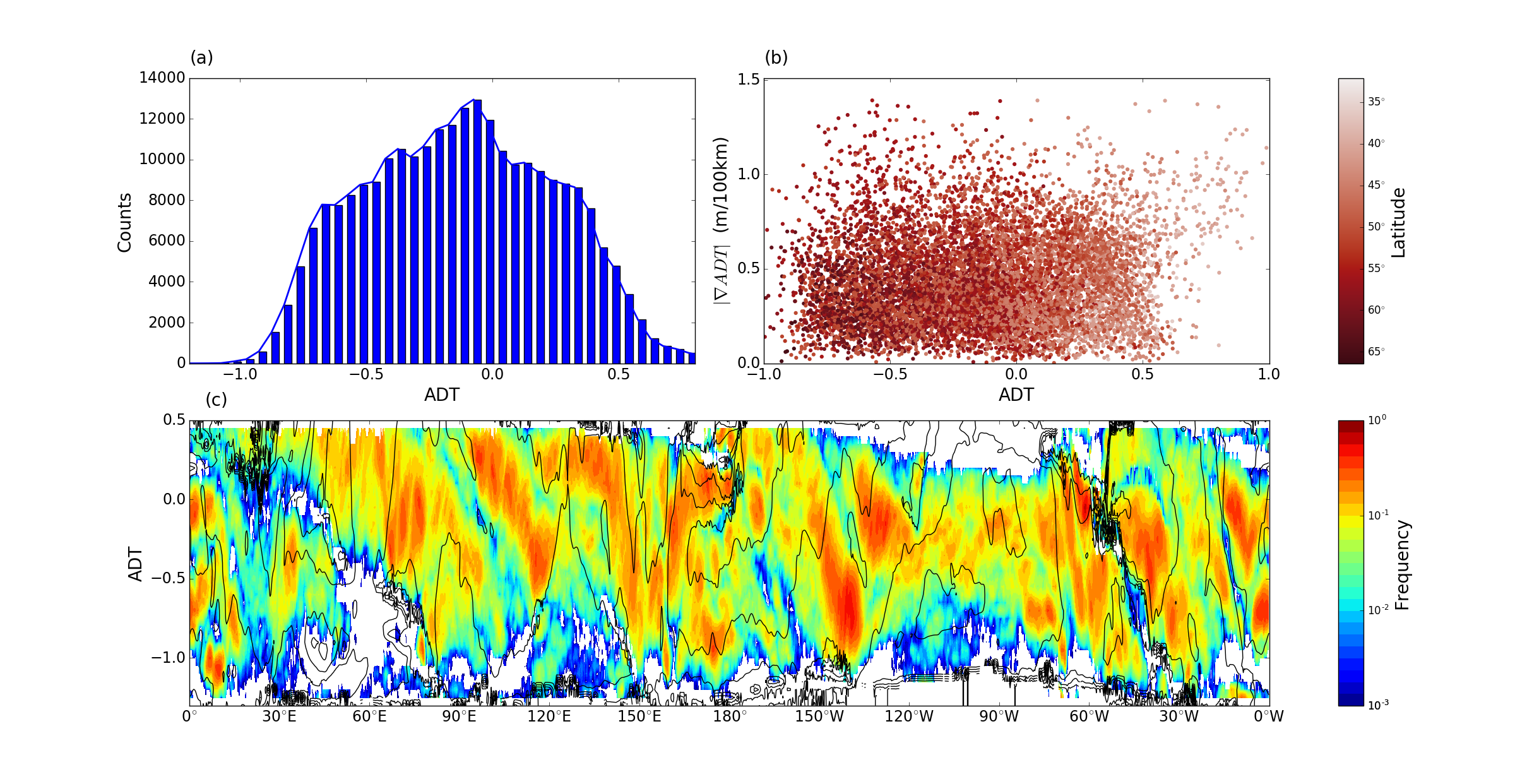}\\
  \caption{The Southern Ocean frontal structure in ADT space. (a) hisogram of the ADT at the frontal locations determined by the WHOSE method; (b) the ADT (x-axis) and the ADT gradient (y-axis) at the frontal locations, colored by latitude; and (c) the frontal frequency of occurance heat-map (as in Fig. \ref{Fig3:Histograms_Centres_Njets}a, in longitude/ADT space. Solid black lines indicate bathymetric contours (CI:1000m) }\label{Fig9:Variability_with_SSH}
\end{figure*}

In order to answer this question, we determine the \textit{time-varying} ADT, as well as the local absolute ADT gradient $|\nabla ADT| = \sqrt{\partial (ADT) / \partial x^{2} + \partial (ADT) / \partial y ^{2}}$ at each of the frontal locations found by the WHOSE method for the entire satellite era. Fig. \ref{Fig9:Variability_with_SSH}a shows the histogram of ADT values for all the fronts detected in Southern Ocean latitudes from 1993 to 2014, which has a similar distribution to the histogram constructed using MDT shown in Fig. \ref{Fig5:Front_Locs_With_MDT}a. Although the ADT histogram is broader than the MDT histogram, the maxima and minima are found at similar ADT values to the MDT values used to determine the frontal labels in section \ref{Section:Mean_Front_Positions}. As such, it is safe to conclude that the ADT can be used to characterize frontal locations in a long-term, broad scale sense.


However, upon further analysis, we find that there is little relationship between the instantaneous position of the front and the corresponding instantaneous location of the an ADT contour. To demonstrate this lack of concordance, we plot the $|\nabla ADT|$ at each front location against the ADT at those same locations (shown in Fig. \ref{Fig9:Variability_with_SSH}b), colored by their latitude to give an idea of their location in physical space. It is evident from Fig. \ref{Fig9:Variability_with_SSH}b that there is no organization of the cloud of points by the value of the ADT. If ADT contours were capable of effectively tracking fronts as they move, we would expect to see a banded structure in Fig. \ref{Fig9:Variability_with_SSH}b, with strong ADT gradients falling on a narrow range of SSH values. However, there is no clear clustering of the frontal locations around any particular value of ADT regardless of the magnitude of its associated ADT gradient (although fronts at lower latitudes are unsurprisingly more likely to have lower values of ADT). 

Finally, the frontal occurrence frequency heat-map of Fig. \ref{Fig3:Histograms_Centres_Njets}a, is recalculated in ADT/longitude coordinates, and shown in Fig. \ref{Fig9:Variability_with_SSH}. Here we see that persistent frontal activity does not necessarily follow ADT contours. In fact, in many locations there is a slow drift across ADT contours. For example, two persistent fronts are present near the Pacific-Antarctic Rise (longitudes 160$^{\circ}$W to 120$^{\circ}$W) that are strongly steered from north to south by the underlying bathymetry (which can be seen in Fig. \ref{Fig3:Histograms_Centres_Njets}a,b). In longitude/ADT space, it is clear that in this region, the fronts drift from approximately -0.1 dyn m to -1.0 dyn m over about 40$^{\circ}$ of longitude -- an approximate distance of 4000km at this latitude -- a decrease in dynamic height of 0.25 dyn m/1000km. While highly persistent fronts (such as those found in regions of strong topographic steering) are generally constrained to a narrow range of latitudes, in longitude/ADT space the regions of high frontal occurance frequencies can be present over a broad range of ADT, and are not necessarily restricted to a particular ADT value. \cite{Thompson&Sallee2012} found similar behavior when analyzing the ADT with running PDFs. In particular, their Fig 2a, which shows a different, yet comparable measure of frontal persistence, has a spatial structure with almost identical form to our Fig. \ref{Fig9:Variability_with_SSH}c.   

We reiterate here that the contour method produces an accurate estimate of the time-mean location of the Southern Ocean front. However, our analysis echoes previous criticism from \cite{GrahamEtAl2012} and \cite{Gille2014} that the time variability of a dynamic height contour may not reflect the time variability of the associated fronts. The complex spatial and temporal variability of the Southern Ocean fronts mean that the results of estimates using contour methods must be interpreted with caution and that attempt to use contour methods to investigate the temporal variability of fronts should be confirmed or validated against other methodologies.

\subsection{Trends in Frontal Location}

\begin{figure}[b!]
   \centering  
  \includegraphics[width=18pc,height=18pc,angle=0]{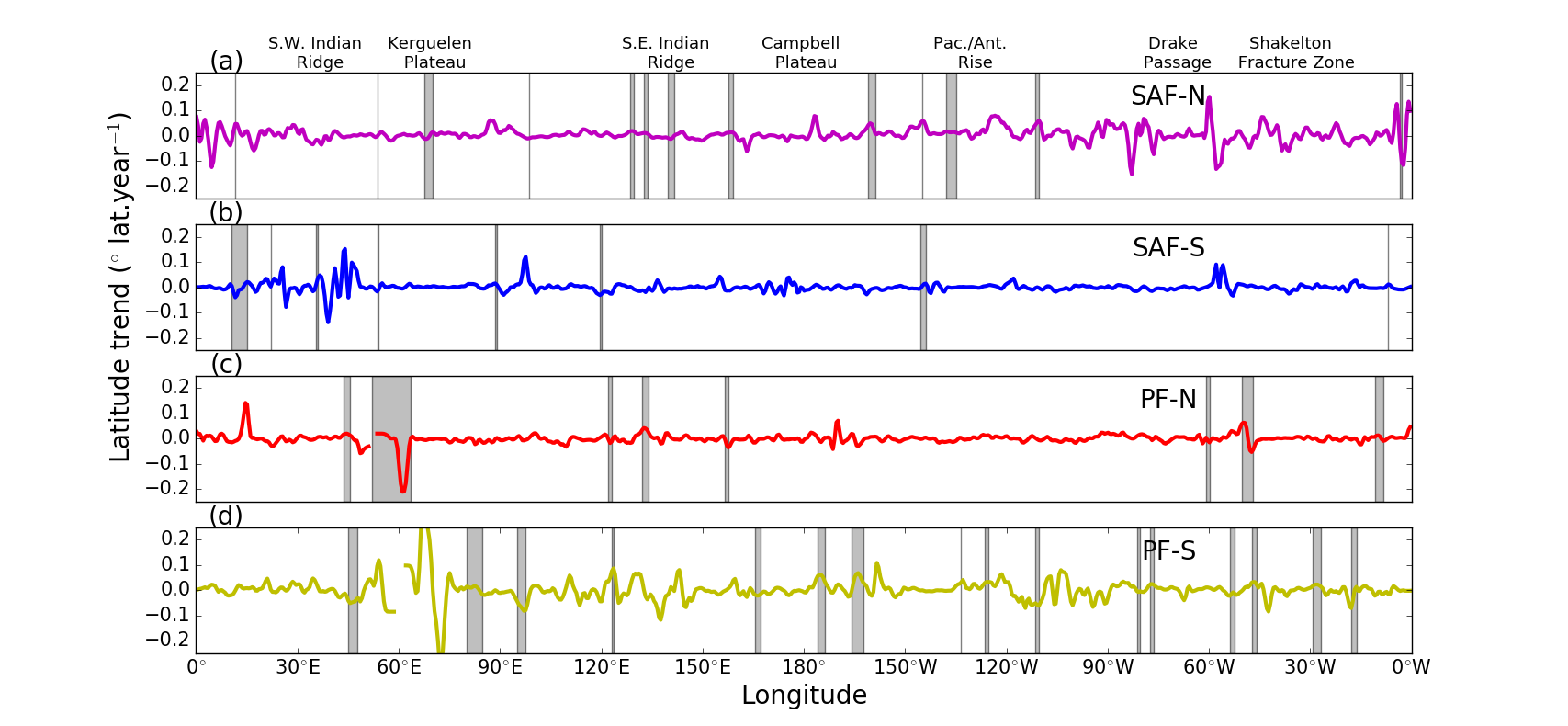}\\
  \caption{Trends in the mean latitudinal position of the (a) SAF-N; (b) SAF-S; (c) PF-N; and (d) PF-S. Grey shading indicates regions where the trend is statistically significant at the 90\% level ($p<0.1$).  }\label{Fig10:Latitude_Trends}
\end{figure}

Using the WHOSE estimates of front, combined with the curve fitting procedure described in section \ref{Section:Mean_Front_Positions} we now estimate the trend in frontal locations. To accomplish this, we form annual heat-maps and fit the skew-normal function, yielding annual-mean frontal locations for each year from 1993 to 2014. We then associated each frontal location with one of the SAF-N, SAF-S, PF-N or PF-S using the MDT values described in section \ref{Section:Mean_Front_Positions} and find the mean latitude for each of the labeled fronts at every longitude and for every year in the database. Trends in the annual mean position of the fronts are then obtained at each longitude using standard linear regression are shown in Fig. \ref{Fig10:Latitude_Trends}.


\begin{figure*}[t!]
   \centering  
  \includegraphics[width=40pc,height=20pc,angle=0]{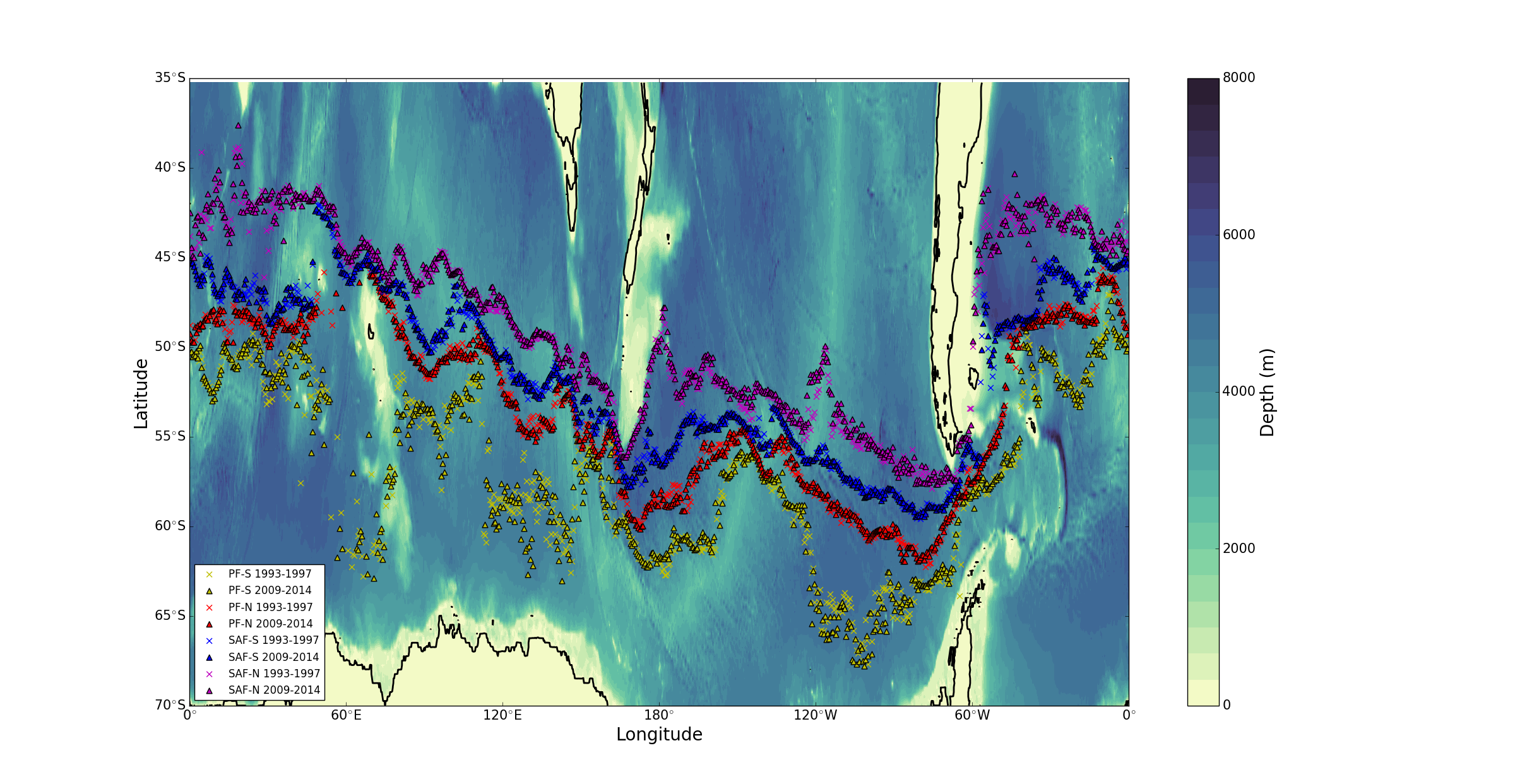}\\
  \caption{Frontal shifts in the Southern Ocean over the altimetric data period. The mean positions of the fronts during the period 1993-1997 ($\times$) and the period 2009-2014 $\bigtriangleup$. The fronts are color coded from north to south as:  SAF-N: magenta, SAF-S: blue; PF-N: red; and PF-S: yellow.  }\label{Fig11:Frontal_Shift}
\end{figure*}

Fig. \ref{Fig10:Latitude_Trends} does not indicate any substantial shifts in the locations of the fronts, although there is some suggestion of a trend of between -0.1$^{\circ}$ and -0.2$^{\circ}$/year in the PF at approximately 60$^{\circ}$E. This region was also found by \cite{Kim&Orsi2014} to host reasonably large (approximately 0.1$^{\circ}$/year) and statistically significant shifts in the PF position. However, despite our $p$-value of less than 0.9 (and it must be emphasized, less than the generally accepted level for statistical significance of $p<$0.05) the lack of spatial coherence in the trend, as well as the lack of any statistically significant trends in the neighboring frontal branches, means that we are unable to place a large degree of confidence in this apparent southward trend being result of any real physical shift in the front's location. To illustrate this point, Fig.  \ref{Fig11:Frontal_Shift} shows the time-mean positions of the fronts for the period 1993-1997 (crosses) and 2009-2014 (triangles). Throughout the Southern Ocean, even in regions where the calculated trend is non-zero and statistically significant, there is no apparent shift in the locations of the fronts.


The lack of trends found in our analysis is contrary to many studies that employ contour type method \citep{SalleeEtAl2008,SokolovAndRintoul2009II,Kim&Orsi2014} that find localized southward shifts of the fronts, between 0.5$^{\circ}$ and 2$^{\circ}$, depending on the details of the methodology, the geographical region and the front under consideration. Our results are, however, consistent with the analyses of \cite{GrahamEtAl2012,Gille2014,ShaoEtAl2015} who find little or no long term trends in the fronts. While it is clear from hydrographic data that the Southern Ocean's sea level is rising \citep{MorrowEtAl20008}, our analysis indicates that it is unlikely to be due to significant deviations in the frontal paths.

\subsection{Response of fronts to ENSO and SAM}

\begin{figure*}[t]
   \centering  
  \includegraphics[width=40pc,height=20pc,angle=0]{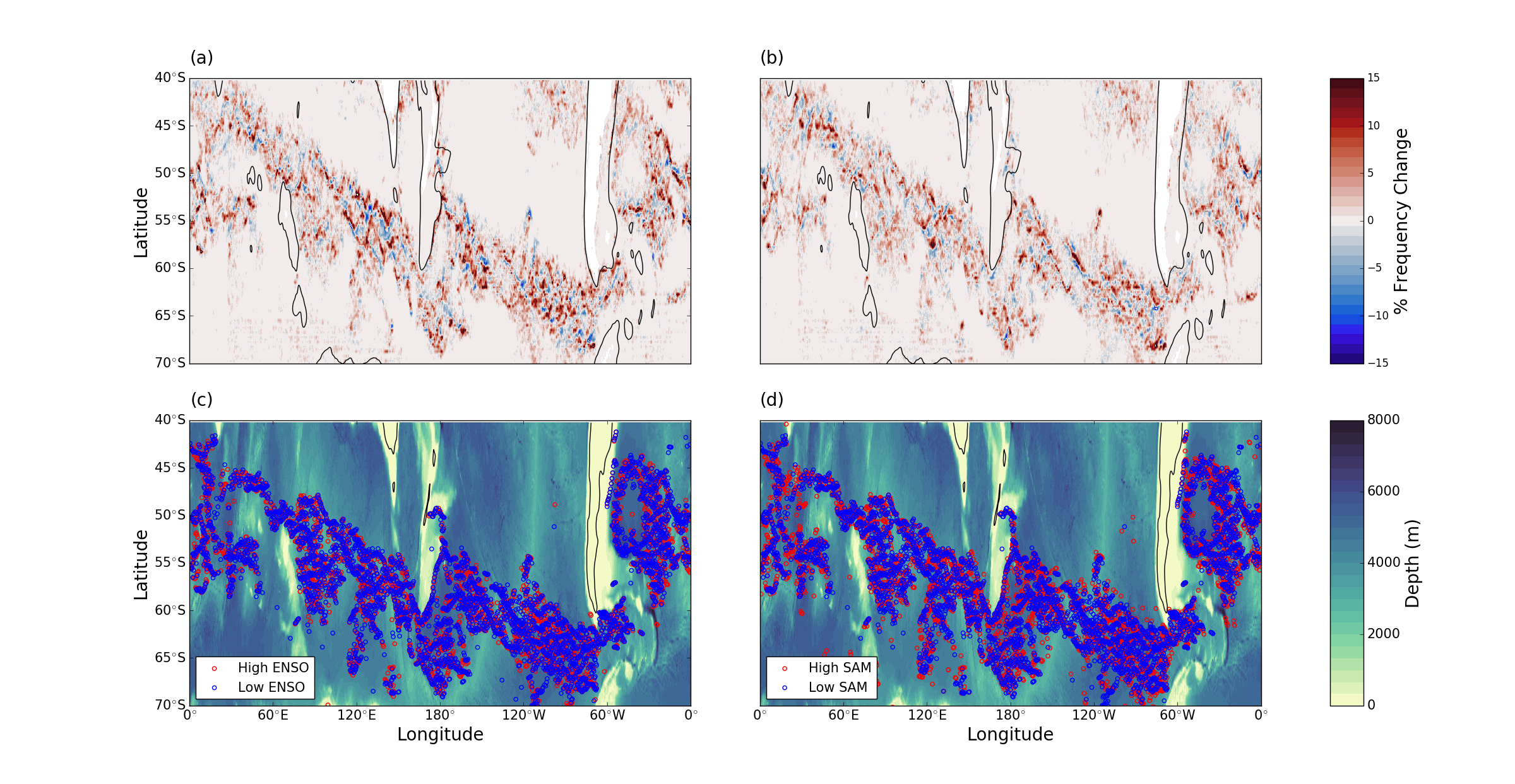}\\
  \caption{The inflence of the SAM and ENSO on frontal positions. (a) The heat-maps anomaly calculated by subtracting the heat-map determind during La Ni\~{n}a periods from El-Ni\~{n}o periods and; (b) as in (a) subtracting the negative SAM heat-maps from the positive SAM heat-maps. `Warm' colors indicate that a front is more likely to be present at that location during positive events, while `cool' colors indicate a tendancy for a front to be present during negative events. The black solid contour shows the coastline. (c) ensemble mean frontal positions during El-Ni\~{n}o (red) and La-Ni\~{n}a periods (blue); and (d)  ensemble mean frontal positions during high SAM events (red) and low SAM events (blue), overlaid over bathymetry.    }\label{Fig12:ENSO_SAM_Difference}
\end{figure*}

Although we have not found any appreciable long term variability, we have not ruled out forced inter--annual variability associated the SAM or ENSO. As in \cite{Kim&Orsi2014}, we designate a `high' (`low') SAM or ENSO event to occur when the value of the index (the \cite{Marshall2003} index for the SAM, and the Bivariate ENSO Timeseries \cite{Smith&Sardeshmukh2000} for ENSO)  was greater (less) than 1 standard deviation. We then form four ensemble-mean frontal occurrence heat--maps: high SAM, low SAM, high ENSO (that is, El Ni\~{n}o) and low ENSO (La Ni\~{n}a). The differences between the ensemble heat-maps calculated during  El Ni\~{n}o and La Ni\~{n}a periods is shown in Fig. \ref{Fig12:ENSO_SAM_Difference}a, while the difference between the heat-maps obtained during strongly positive and strongly negative SAM periods is shown in Fig. \ref{Fig12:ENSO_SAM_Difference}b.


For both climate modes, we find differences local changes in the frontal occurrence frequency of as much as 15\%. There is also some indication of coherent frontal shifts forced by atmospheric variability in certain regions. For example, downstream of the Kerguelen Plateau (at approximately 110$^{\circ}$E), south of Australia  (130$^{\circ}$E) and in Drake Passage (60$^{\circ}W$), there is evidence of a frontal shift to the south during La Ni\~{n}a periods. Similarly, in the eastern Pacific, near the Pacific--Antarctic Rise (140$^{\circ}$W) there is evidence of frontal shifts to the south during negative SAM periods. These shifts are, however, small: less than 1.0$^{\circ}$ of latitude and generally less than 0.5$^{\circ}$ which is approaching the effective resolution of the gridded data-set.  

To further underscore the weak dependence of the frontal structure on the climate modes, we plot in Fig. \ref{Fig12:ENSO_SAM_Difference}b,c the time-mean frontal locations during high and low ENSO or SAM events, determined using the skew-normal curve fitting procedure. As before, we find no clear evidence of substantial frontal shifts with changes in either climate mode. It is important to note, just as in the model output analyzed by \cite{GrahamEtAl2012}, there appears to be no substantial movement of the fronts over flat bottomed regions such as the east Pacific. The largest movements of the fronts appears to be along the northern boundary of the ACC, although these shifts are neither spatially coherent or large in magnitude.

Essentially, we have not found significant sensitivity of fronts to climate modes, although there is the suggestion of limited variability in a few select regions. As with the analysis of long term trends, our results are at odds with a number of studies employing contour methods who show localized sensitivity to ENSO or SAM \citep{SalleeEtAl2008,Kim&Orsi2014}, yet in broad agreement with other who find limited or no response to changes in atmospheric forcing \citep{GrahamEtAl2012,ShaoEtAl2015}, although it is notable that the later study does find statistically significant correlation between frontal position and the SAM in a few limited geographic regions (although these correlations do not imply substantial displacements in the frontal positions). Any shifts in frontal position detected here are generally limited to less than 1.0$^{\circ}$ of latitude, and have a coherent spatial extent of no more than 10$^{\circ}$ of longitude.

\section{Discussion and Conclusion} \label{Conclusion}

In this paper we have applied the WHOSE method, described in \cite{Chapman2014} to 21 years of gridded sea-surface height altimetry in order to investigate the spatial and temporal variability of meso-scale fronts in the Southern Ocean. We have attempted to tease out concrete conclusions from a complex dataset by forming `heat-maps' of the frontal occurrence frequency and then approximating these maps by fitting a superposition of skew-normal curves. This approach allows us to quasi-objectively determine the number of fronts at each latitude and their time-mean position. With these data, we are able to investigate in detail their spatial and temporal variability of the frontal structure. We have also compared our methodology with the commonly used `contour methods' and attempted to compare the utility of the two approaches.  

The principle result of this study is the identification of marked spatial variation in the frontal structure throughout the Southern Ocean latitudes. We find that downstream of large bathymetric features and over regions of relatively flat bottom topography, the frontal structure responds by `splitting' into a number of sub-fronts, whereas over large bathymetry, the fronts are constrained are `merge' together. This splitting/merging variability is distinct from meandering, where a front changes its location yet remains coherent. In several regions within the Southern Ocean splitting behaviour dominates over the meandering of fronts. The distinction is important, as splitting represents a rearrangement of the frontal structure, which in turn indicates a rearrangement of the potential vorticity structure \citep{ThompsonEtAl2010,Thompson&Sallee2012} in a way that a simple meandering of a front does not. 

The second key results arising from this work is that the fronts determined by our methodology show no significant trends in their position and only limited, and localized, sensitivity to the changes in atmospheric forcing associated with the Southern Annual Mode or El Ni\~{n}o-Southern Oscillation. The lack of temporal variability is consistent with the results of previous studies \citep{GrahamEtAl2012,Gille2014,ShaoEtAl2015}. There are, however, several studies employing contour methods that have shown trends in the position of the fronts in particular locations \citep{SokolovAndRintoul2009II,Kim&Orsi2014}. In contrast our results do not indicate coherent trends in any region, with the possible exception of in the southeast Indian sector. 

We have sought to identify the reasons for the discrepancy between methodologies by investigating how well contours of dynamic sea-surface topography approximate the frontal locations determined by the WHOSE method. It was shown in Sec. \ref{Section:Mean_Front_Positions} that contour methods do accurately represent the time-mean locations and, for the most-part, orientations, of the WHOSE fronts. However, when considering the time variability of fronts, contour methods have several short-comings. Firstly, as shown in Fig. \ref{Fig9:Variability_with_SSH}b, and echoing the work of \cite{GrahamEtAl2012}, high dynamic topography gradients are not always clustered around a restricted set of sea-surface height values. Secondly, and perhaps more importantly, the reorganization of the frontal structure downstream of large bathymetry, leads to fronts changing their expression in SSH space. In certain regions, such as near the Pacific-Antarctic Rise in the east Pacific, fronts cross or drift across SSH contours downstream of a large topographic feature 

Despite the shortcomings of contour methods, all is not lost and these methods can still remain invaluable due to their simplicity of application, relationship with traditional hydrographic fronts and accurate representation of time mean frontal positions, particularly as a `circumpolar coordinate system' \citep{LanglaisEtAl2011}. However, we make the following recommendations to researchers using these methods in future work:
\begin{enumerate}
\item when attempting to use contour methods to characterize the variability of frontal location, be particularly careful when interpreting the results and, where possible, validate any frontal movement with another method or with independent data, such as hydrography; and
\item be mindful of the fact that the reorganization of the frontal structure may impact the local association between the fronts and SSH. It may be more accurate to fit SSH contours to front locations in limited geographic ranges.
\end{enumerate}

Future work will focus on investigating the implications that the continual reorganization of the frontal structure has for the large scale circulation in the Southern Ocean, as well as more rigorously and precisely defining the frontal `regimes' present. 

\section*{Acknowledgments} 
This research was funded by a National Science Foundation Division of Ocean Sciences Postdoctoral Fellowship number 1521508. The author thanks Robert Graham and Sabine Arnault for comments on an earlier draft of this manuscript and Jean-Baptiste Sall\'{e}e for useful discussions and encouragement.

\section*{Appendix A: Software Availability}
Implementations of the WHOSE method and the curve fitting procedure, written in the open source Python programming language, is available as open source software (under an MIT license) from the author's GitHub repository: https://github.com/ChrisC28

\section*{Appendix B: Availability of WHOSE front locations}
The frontal locations determined by the WHOSE method have been made freely available for download at the following URL: 
The data is stored in NetCDF files, and consist of daily output between latitudes 70$^{\circ}$S and 35$^{\circ}$S, with a grid spacing of 1/4$^{\circ}$, identical to that of the AVISO data from which it is generated. The data are binary: for each point in the domain, a `TRUE' value indicates the presence of a front, while a FLASE indicates that no front was found at that location or time.



\end{document}